\documentclass[11pt,a4paper]{article}
\usepackage[margin=2.5cm]{geometry}
\usepackage[T1]{fontenc}
\usepackage[utf8]{inputenc}
\usepackage{lmodern}
\usepackage{amsmath,amssymb,amsfonts,mathtools}
\numberwithin{equation}{section}
\usepackage{booktabs,array,longtable}
\usepackage{graphicx}
\usepackage{float}
\usepackage{hyperref}
\hypersetup{colorlinks=true,linkcolor=black,citecolor=black,urlcolor=blue}
\usepackage{microtype}
\usepackage{enumitem}
\usepackage{authblk}
\usepackage{xcolor}

\newcommand{\CS}{\operatorname{CS}}
\newcommand{\so}{\mathfrak{so}}
\newcommand{\iso}{\mathfrak{iso}}

\begin{document}

\title{Resonant JPZ algebras and 3D Chern--Simons gravity}

\date{}

\setcounter{Maxaffil}{0}
\renewcommand\Affilfont{\itshape\small}
\renewcommand\Authfont{\normalfont}

\author[1]{Remigiusz Durka%
\thanks{remigiusz.durka@uwr.edu.pl}}

\author[2,3]{Nelson Merino%
\thanks{nemerino@unap.cl}}

\author[2,3]{Diego Molina Peñafiel%
\thanks{dimolina@unap.cl}}

\affil[1]{Institute for Theoretical Physics, University of Wroc\l{}aw,\protect\\ pl.\ M.\ Borna 9, 50-204 Wroc\l{}aw, Poland}

\affil[2]{Instituto de Ciencias Exactas y Naturales,\protect\\ Universidad Arturo Prat, Playa Brava 3265, 1111346, Iquique, Chile}

\affil[3]{Facultad de Ciencias, Universidad Arturo Prat,\protect\\ Avenida Arturo Prat Chacón 2120,
1110939, Iquique, Chile}

\maketitle

\begin{abstract}
In this work we construct three-dimensional Chern--Simons gravity theories for the six resonant $JPZ$ algebras generated by the Lorentz generators $J_a$, translations $P_a$, and an additional Lorentz-like generator $Z_a$. The resonance condition together with the Jacobi identities leaves precisely six configurations arising from the semigroup expansions. Our main result is that three of them, $B_4$, $\widetilde C_4$, and $C_4$, are isomorphic to $\iso(2,1)\oplus\so(2,1)$, although they yield different gravitational realizations when we keep the original $JPZ$ field identification fixed. For each case $JPZ$, we determine the invariant tensors, gauge curvatures, Chern--Simons action, field equations, and the corresponding maximal-rank coupling sectors. The field equations can be written in the compact form $\Omega_\sigma\mathbf F=0$, which makes explicit how the additional field $h^a$ contributes to the curvature, torsion, and the remaining gauge-curvature components. We use the direct-sum isomorphisms to classify the local metric sectors and to relate them to known global and asymptotic solutions, while keeping the gravitational interpretation tied to the original $(\omega,e,h)$ fields.
\end{abstract}


\section{Introduction}
\label{sec:introduction}

Three-dimensional gravity provides a natural setting to study the relation between gauge symmetry, invariant tensors, and gravitational variables. Chern--Simons gravity turns the choice of Lie algebra and invariant bilinear form into a dynamical framework, while identifying particular one-forms as the spin connection and dreibein gives an additional gravitational interpretation \cite{Achucarro:1986uwr,Witten:1988hc,Hassaine:2016amq}.

A systematic way to obtain enlarged symmetries is through Lie-algebra enlargements, in particular, through the $S$-expansion \cite{Izaurieta:2006zz,Edelstein:2006se,Izaurieta:2009hz,Andrianopoli:2013ooa,Diaz:2012zza,Inostroza:2018gzd} and resonant-algebra constructions \cite{Durka:2016eun,Durka:2019vnb,Durka:2021glk}. For the AdS decomposition $\mathfrak g=V_0\oplus V_1$, with $[V_0,V_0]\subset V_0$, $[V_0,V_1]\subset V_1$, and $[V_1,V_1]\subset V_0$, the resonance condition is implemented by a semigroup decomposition $S=S_0\cup S_1$ satisfying $S_0S_0\subset S_0$, $S_0S_1\subset S_1$, and $S_1S_1\subset S_0$. Resonant constructions preserve a prescribed decomposition pattern of the original (AdS) algebra and have been used to organize large families of gravitational and supersymmetric extensions \cite{Durka:2022nih,Durka:2022ghv,Concha:2020atg}. Following previous resonant constructions, we use the AdS algebra as the parent algebra of the expansion. This choice is motivated by possible supergravity applications. 

In the present work, we focus on a three-generator content
\begin{equation}
\{J_a,P_a,Z_a\},
\end{equation}
where $J_a$ spans the Lorentz subalgebra, $P_a$ is a translation generator, and $Z_a$ is an additional Lorentz vector. Imposing the resonant multiplication pattern together with the Jacobi identities leaves precisely six admissible configurations \cite{Durka:2016eun,Durka:2019vnb,Durka:2021glk,Durka:2022ghv},
\begin{equation}
\widetilde B_4,\qquad B_4,\qquad \widetilde C_4,\qquad C_4,\qquad \mathfrak B_4,\qquad \mathfrak C_4\,.
\end{equation}

Several members of this list are already known and have been studied extensively. Maxwell-type originate in the symmetry of particles in constant electromagnetic backgrounds and have subsequently been studied through contractions, deformations, expansions, gravity, and supergravity \cite{Bacry:1970ye,Schrader:1972zd,Bonanos:2010fw,deAzcarraga:2010sw,Lukierski:2010dy,Kamimura:2011mq,deAzcarraga:2012zv,Concha:2014tca,Concha:2015woa,Concha:2016hbt,Concha:2019icz,Salgado-Rebolledo:2019kft}. Their three-dimensional Chern--Simons realization has a non-trivial asymptotic and boundary structure \cite{Concha:2018zeb,Concha:2024rac,Hofenstock:2025xoc}. The semisimple enlargement $\mathfrak{C}_4$, often referred to as the Soroka--Soroka or AdS--Lorentz algebra \cite{Durka:2011nf,Durka:2011gm}, provides the complementary branch and admits direct-sum descriptions and Chern--Simons realizations \cite{Soroka:2006aj,Soroka:2011tc,Salgado:2013eut,Fierro:2014lka,Concha:2018jjj}. Related developments on expansion methods, invariant tensors, and three-dimensional Chern--Simons actions appear in Refs.~\cite{Andrianopoli:2013ooa,Diaz:2012zza,Inostroza:2018gzd,Barriga:2026awj}. Recent Maxwell constructions also show how the same expansion logic organizes non- and ultra-relativistic kinematical gravities with non-degenerate invariant forms \cite{Concha:2026vmg}. Complementary developments within the Mielke--Baekler framework include non-relativistic and Newtonian Chern--Simons gravity, as well as Carrollian gravity with torsion \cite{Concha:2023ejs,Concha:2025vhd}.

A central result of this work is the discovery that three of the six resonant $JPZ$ configurations belong to the same abstract isomorphism class,
\begin{equation}
B_4\simeq\widetilde C_4\simeq C_4
\simeq{\text{Poincar\'e}}\oplus\text{Lorentz}\,.
\label{eq:intro-triple-isomorphism}
\end{equation}
This threefold identification is relevant as the original $JPZ$ embedding differs across the three configurations. Throughout the paper, by the \emph{Lorentz embedding} we mean the fixed identification of $J_a$ as the Lorentz subalgebra, with $P_a$ and $Z_a$ transforming as Lorentz vectors. We identify the spin connection $\omega^a$ as the gauge field associated with the distinguished here generator $J_a$, while $e^a$ is the dreibein and $h^a$ is the additional one-form. A direct-sum basis may mix these fields and is extremely useful computationally, but every geometric statement must be translated back to the original variables. In particular, $\omega$ need not coincide with the Levi--Civita connection of the metric built from $e$; whenever a shifted combination such as $\omega+h$ becomes the torsionless metric connection, this follows from the field equations rather than from redefining the original Lorentz embedding. 

At the dynamical level, for all six algebra configurations we determine the invariant bilinear form, gauge curvatures, Chern--Simons Lagrangian, and bulk field equations. The variational problem takes the universal matrix form. This form separates the curvature vector $\mathbf F$ (composed out of the curvatures and fixed by the algebra) from the combinations of its components selected by the invariant tensor $\Omega_\sigma$. Our main dynamical result is that, in the non-degenerate sectors, the equations imply \emph{full gauge flatness}, by which we mean $F_J^a=F_P^a=F_Z^a=0$, i.e., all $JPZ$  gauge curvatures vanish. Among the six resonant cases, the same additional field $h^a$ enters different geometrical sectors: depending on the algebra, it contributes to Lorentz curvature, torsion, its own curvature, or to the $e\times e$ sector. The resulting regular bulk systems therefore directly illustrate how a single extra gauge field can, through the specific algebra, contribute to different parts of three-dimensional geometry.

The direct-sum viewpoint then makes the local solution structure transparent. The $B_4$, $\widetilde C_4$, $C_4$, and $\mathfrak B_4$ metric sectors are locally flat, while $\mathfrak C_4$ yields locally AdS$_3$ geometry; $\widetilde B_4$ is intrinsically rank deficient. This local statement is separated from global and boundary information. 

The paper is organized as follows. Section~\ref{sec:jpz-configurations}
introduces the six $JPZ$ configurations and their invariant tensors.
Sections~\ref{sec:lagrangian-content} and~\ref{sec:bulk-dynamics} compare
the resulting Chern--Simons actions and bulk equations. The direct-sum
isomorphisms are discussed in Section~\ref{sec:isomorphisms}, and the local
solutions are obtained in Section~\ref{sec:local-solutions-v3}. Their global and
asymptotic interpretation is given in
Section~\ref{sec:standard-geometries-v3}, followed by some further comments concerning $C_4$. The appendices collect the explicit Lagrangians and variational equations, the degenerate coupling sectors, and
the relation with standard Poincar\'e, (A)dS, and Mielke--Baekler gravity.

\section{Six Lorentz-embedded JPZ configurations}
\label{sec:jpz-configurations}

We work on a three-dimensional manifold $\mathcal M$ with spacetime indices denoted by $\mu,\nu,\rho,\ldots=0,1,2$. Lorentz indices are denoted by $a,b,c,\ldots=0,1,2$. The tangent-space Minkowski metric and the Levi-Civita symbol are fixed as
\begin{equation}
\eta_{ab}=\mathrm{diag}(-,+,+)\,,\qquad \epsilon_{012}=+1\,.
\end{equation}
Whenever a spacetime metric interpretation is required, we use
\begin{equation}
g_{\mu\nu}=\eta_{ab}e^a{}_{\mu}e^b{}_{\nu}\,.
\label{eq:metric-definition}
\end{equation}

In three dimensions, antisymmetric Lorentz generators can be dualized as $J_a=\frac12\epsilon_a{}^{bc}J_{bc}$, so that
\begin{equation}
[J_a,J_b]=\epsilon_{ab}{}^cJ_c\,.
\label{eq:lorentz-algebra}
\end{equation}
The gauge connection associated with the Lorentz generator $J_a$ is denoted by $\omega^a$, and its curvature two-form is
\begin{equation}
R^a(\omega)=d\omega^a+\frac12\epsilon^a{}_{bc}\omega^b\wedge\omega^c\,.
\label{eq:lorentz-curvature}
\end{equation}
For $p$-form $v^a$, the Lorentz-covariant derivative is
\begin{equation}
D_\omega v^a=dv^a+\epsilon^a{}_{bc}\omega^b\wedge v^c\,.
\label{eq:lorentz-covariant-derivative}
\end{equation}
In particular, the torsion associated with a dreibein $e^a$ is
\begin{equation}
T^a=D_\omega e^a=de^a+\epsilon^a{}_{bc}\omega^b\wedge e^c\,.
\label{eq:torsion-definition}
\end{equation}

All the algebraic configurations are generated by $\{J_a,P_a,Z_a\}$, with $J_a$ spanning the Lorentz subalgebra. The remaining generators transform according to
\begin{equation}
[J_a,P_b]=\epsilon_{ab}{}^cP_c\,,\qquad [J_a,Z_b]=\epsilon_{ab}{}^cZ_c\,.
\label{eq:lorentz-action}
\end{equation}

We call this fixed identification of $J_a$ as the Lorentz subalgebra, with $P_a$ and $Z_a$ transforming as Lorentz vectors, the Lorentz embedding. This is essential for interpreting the physical spin connection.

The six $JPZ$ cases \cite{Durka:2016eun,Durka:2019vnb,Durka:2021glk,Durka:2022ghv} share commutators~\eqref{eq:lorentz-algebra} and \eqref{eq:lorentz-action}, and they differ only in $[P,P]$, $[P,Z]$, and $[Z,Z]$. One can see the differences in the table, where we put entries always multiplied by $\epsilon_{ab}{}^c$, so $[X_a,Y_b]=\epsilon_{ab}{}^c(\text{entry})_c$. 
\begin{table}[H]
\centering
\renewcommand{\arraystretch}{1.2}
\begin{tabular}{c||c|c|c}
Configuration & $[P_a,P_b]$ & $[P_a,Z_b]$ & $[Z_a,Z_b]$\\
\midrule
\midrule
$\widetilde B_4$ & $0$ & $0$ & $0$\\
$B_4$ & $0$ & $P_c$ & $J_c$\\
$\widetilde C_4$ & $0$ & $0$ & $Z_c$\\
$C_4$ & $0$ & $P_c$ & $Z_c$\\
$\mathfrak B_4$ & $Z_c$ & $0$ & $0$\\
$\mathfrak C_4$ & $Z_c$ & $P_c$ & $Z_c$\\
\end{tabular}
\caption{The six Lorentz-embedded $JPZ$ configurations.}
\label{tab:jpz-configurations}
\end{table}
The first four cases above are Poincar\'e-like (in the algebraic sense, they obey $[P_a, P_b]=0$), while the last two are Maxwell-like, with $[P, P]\sim Z$. No member of the resonant $JPZ$ family realizes $[P, P]\sim J$. 
\subsection{Invariant tensors and coupling spaces}
\label{sec:invariant-tensors}

The commutation relations between generators alone do not determine the Chern--Simons theory. A symmetric invariant bilinear form is also required. All configurations share
\begin{equation}
\langle J_a,J_b\rangle=\alpha_0\eta_{ab}\,,\quad
\langle J_a,P_b\rangle=\alpha_1\eta_{ab}\,,\quad
\langle J_a,Z_b\rangle=\alpha_2\eta_{ab}\,,
\end{equation}
where $\alpha_0$, $\alpha_1$, and $\alpha_2$ are arbitrary real parameters.
The remaining entries are shown below:

\begin{table}[ht]
\centering
\renewcommand{\arraystretch}{1.2}
\begin{tabular}{c||c|c|c}
Configuration & $\langle P,P\rangle/\eta$ & $\langle P,Z\rangle/\eta$ & $\langle Z,Z\rangle/\eta$\\
\midrule
\midrule
$\widetilde B_4$ & $0$ & $0$ & $0$\\
$B_4$ & $0$ & $\alpha_1$ & $\alpha_0$\\
$\widetilde C_4$ & $0$ & $0$ & $\alpha_2$\\
$C_4$ & $0$ & $\alpha_1$ & $\alpha_2$\\
$\mathfrak B_4$ & $\alpha_2$ & $0$ & $0$\\
$\mathfrak C_4$ & $\alpha_2$ & $\alpha_1$ & $\alpha_2$\\
\end{tabular}
\caption{Configuration-dependent invariant-tensor components.}
\label{tab:invariant-components}
\end{table}
The invariant-tensor components~\cite{Durka:2016eun,Concha:2016hbt} can be obtained directly from the $S$-expansion procedure \cite{Izaurieta:2006zz}.

In the ordered basis $(J,P,Z)$ we can write the bilinear form as $\Omega_\sigma\otimes\eta_{ab}$, where $\sigma$ labels the algebraic configuration. After suppressing the common $\eta_{ab}$ factor, the corresponding $\Omega$ matrices are
\begin{align}
\Omega_{\widetilde B_4}&=\begin{pmatrix}\alpha_0&\alpha_1&\alpha_2\\\alpha_1&0&0\\\alpha_2&0&0\end{pmatrix},&
\Omega_{B_4}&=\begin{pmatrix}\alpha_0&\alpha_1&\alpha_2\\\alpha_1&0&\alpha_1\\\alpha_2&\alpha_1&\alpha_0\end{pmatrix},\notag\\
\Omega_{\widetilde C_4}&=\begin{pmatrix}\alpha_0&\alpha_1&\alpha_2\\\alpha_1&0&0\\\alpha_2&0&\alpha_2\end{pmatrix},&
\Omega_{C_4}&=\begin{pmatrix}\alpha_0&\alpha_1&\alpha_2\\\alpha_1&0&\alpha_1\\\alpha_2&\alpha_1&\alpha_2\end{pmatrix},\notag\\
\Omega_{\mathfrak B_4}&=\begin{pmatrix}\alpha_0&\alpha_1&\alpha_2\\\alpha_1&\alpha_2&0\\\alpha_2&0&0\end{pmatrix},&
\Omega_{\mathfrak C_4}&=\begin{pmatrix}\alpha_0&\alpha_1&\alpha_2\\\alpha_1&\alpha_2&\alpha_1\\\alpha_2&\alpha_1&\alpha_2\end{pmatrix}\,.\label{matrices}
\end{align}
The determinant of $\Omega_{\widetilde B_4}$ vanishes identically, so this case is intrinsically degenerate, whereas each remaining configuration has a non-degenerate sector whenever the corresponding determinant is non-zero.

\section{Chern--Simons Lagrangians}
\label{sec:lagrangian-content}

\subsection{Gauge connection and curvatures}
\label{sec:gauge-connection}

The gauge connection one-form valued in each of the six $JPZ$ algebras is given by
\begin{equation}
\mathcal A=\omega^a J_a+\frac{1}{\ell}e^a P_a+h^a Z_a\,,
\label{eq:connection_def}
\end{equation}
where $\omega^a$ is the Lorentz spin connection associated with $J_a$, $e^a$ is the dreibein, and $h^a$ is the additional one-form. The scale $\ell$ is a non-vanishing length introduced to fix dimensionality. It becomes the actual (A)dS radius only in the regular $\mathfrak C_4$ metric sector.

The gauge curvature is
\begin{equation}
\mathcal F=d\mathcal A+\frac12[\mathcal A,\mathcal A]=F_J^a J_a+F_P^a P_a+F_Z^a Z_a\,,
\label{eq:gauge-curvature}
\end{equation}
where $F_J^a$, $F_P^a$, and $F_Z^a$ denote the curvature components along the generators $J_a$, $P_a$, and $Z_a$, respectively. Their geometrical content depends on the underlying algebra as shown in Table~\ref{tab:curvatures}.

\begin{table}[ht]
\centering
\small
\renewcommand{\arraystretch}{1.3}
\begin{tabular}{c||c|c|c}
Configuration & $F_J^a$ & $F_P^a$ & $F_Z^a$\\
\midrule
\midrule
$\widetilde B_4$ & $R^a$ & $\ell^{-1}T^a$ & $D_\omega h^a$\\
$B_4$ & $R^a+\frac12(h\times h)^a$ & $\ell^{-1}(T^a+(h\times e)^a)$ & $D_\omega h^a$\\
$\widetilde C_4$ & $R^a$ & $\ell^{-1}T^a$ & $D_\omega h^a+\frac12(h\times h)^a$\\
$C_4$ & $R^a$ & $\ell^{-1}(T^a+(h\times e)^a)$ & $D_\omega h^a+\frac12(h\times h)^a$\\
$\mathfrak B_4$ & $R^a$ & $\ell^{-1}T^a$ & $D_\omega h^a+\frac{1}{2\ell^2}(e\times e)^a$\\
$\mathfrak C_4$ & $R^a$ & $\ell^{-1}(T^a+(h\times e)^a)$ & $D_\omega h^a+\frac{1}{2}(h\times h)^a+\frac{1}{2\ell^2}(e\times e)^a$\\
\end{tabular}
\caption{Gauge-curvature components.}
\label{tab:curvatures}
\end{table}

Here we use $(e\times e)^a=\epsilon^a{}_{bc}e^b\wedge e^c$ and $(h\times h)^a=\epsilon^a{}_{bc}h^b\wedge h^c$, as well as $(h\times e)^a=\epsilon^a{}_{bc}h^b\wedge e^c$.

The gauge transformations are generated by $\delta_\Lambda\mathcal A=d\Lambda+[\mathcal A,\Lambda]$, with 
\begin{equation}
\Lambda=\lambda^aJ_a+\ell^{-1}\tau^aP_a+\chi^aZ_a\,.
\label{eq:gauge_parameter}
\end{equation}
Under an infinitesimal Lorentz transformation with parameter $\rho^a$,
\begin{align}
\delta_\lambda\omega^a&=D_\omega\rho^a\,,\notag\\
\delta_\lambda e^a&=\epsilon^a{}_{bc}e^b\rho^c\,,\notag\\
\delta_\lambda h^a&=\epsilon^a{}_{bc}h^b\rho^c\,.
\end{align}

The curvature transforms covariantly, $\delta_\Lambda\mathcal F=[\mathcal F,\Lambda]$, and satisfies the Bianchi identity $D_{\mathcal A}\mathcal F=d\mathcal F+[\mathcal A,\mathcal F]=0$.

\subsection{Chern--Simons actions}
\label{sec:cs-actions}

The Chern--Simons action is
\begin{equation}
I_\sigma[\mathcal A]=\frac{k}{4\pi}\int_{\mathcal M}\left\langle\mathcal A\wedge d\mathcal A+\frac13\mathcal A\wedge[\mathcal A,\mathcal A]\right\rangle_\sigma=\frac{k}{4\pi}\int_{\mathcal M} \mathcal L_\sigma\,.
\label{eq:cs-action}
\end{equation}
Up to an exact form, the Lagrangian density can be written as
\begin{align}
\mathcal L_\sigma={}&
\alpha_0\CS(\omega)
+\frac{2\alpha_1}{\ell}e_a\wedge R^a
+2\alpha_2h_a\wedge R^a
\nonumber\\
&+\frac{\beta_{PP}}{\ell^2}e_a\wedge D_\omega e^a
+\frac{2\beta_{PZ}}{\ell}e_a\wedge D_\omega h^a
+\beta_{ZZ}h_a\wedge D_\omega h^a
\nonumber\\
&+\frac{\gamma_{PPP}}{3\ell^3}
\epsilon_{abc}e^a\wedge e^b\wedge e^c
+\frac{\gamma_{PPZ}}{\ell^2}
\epsilon_{abc}e^a\wedge e^b\wedge h^c
\nonumber\\
&+\frac{\gamma_{PZZ}}{\ell}
\epsilon_{abc}e^a\wedge h^b\wedge h^c
+\frac{\gamma_{ZZZ}}{3}
\epsilon_{abc}h^a\wedge h^b\wedge h^c\,,
\label{eq:universal-cs-density}
\end{align}
where the Lorentz Chern--Simons three-form is
\begin{equation}
\CS(\omega)=\omega_a\wedge d\omega^a+\frac13\epsilon_{abc}\omega^a\wedge\omega^b\wedge\omega^c\,.
\end{equation}
For each algebra, the coefficients of the invariant tensors are as follows:
\begin{align*}
\widetilde B_4 &: \quad \text{all additional $\beta$- and $\gamma$-coefficients vanish}\,,\\
B_4 &: \quad \beta_{PZ}=\gamma_{PZZ}=\alpha_1,\qquad \beta_{ZZ}=\alpha_0,\qquad \gamma_{ZZZ}=\alpha_2\,,\\
\widetilde C_4 &: \quad \beta_{ZZ}=\gamma_{ZZZ}=\alpha_2\,,\\
C_4 &: \quad \beta_{PZ}=\gamma_{PZZ}=\alpha_1,\qquad \beta_{ZZ}=\gamma_{ZZZ}=\alpha_2\,,\\
\mathfrak B_4 &: \quad \beta_{PP}=\alpha_2\,,\\
\mathfrak C_4 &: \quad \beta_{PP}=\gamma_{PPZ}=\beta_{ZZ}=\gamma_{ZZZ}=\alpha_2,\qquad
\beta_{PZ}=\gamma_{PZZ}=\gamma_{PPP}=\alpha_1\,.
\end{align*}
The coefficients not listed here vanish.

For convenience, we collect and compare the term \emph{content} of the six actions below, hiding the wedge by $e\cdot R = e^a \wedge R_a$ and the epsilon by the $\times$ symbol. We also collect the individual Lagrangians with all explicit indices in Appendix~\ref{app:explicit-actions}.

\begin{table}[!ht]
\centering
\small
\setlength{\tabcolsep}{8pt}
\renewcommand{\arraystretch}{1.55}
\begin{tabular}{l||cccccc}
Lagrangian
& $\widetilde B_4$ & $B_4$ & $\widetilde C_4$ & $C_4$
& $\mathfrak B_4$ & $\mathfrak C_4$\\
\midrule
\midrule
$2\ell^{-1}e\cdot R$
& $\alpha_1$ & $\alpha_1$ & $\alpha_1$ & $\alpha_1$ & $\alpha_1$ & $\alpha_1$\\
$2\ell^{-1}e\cdot D_\omega h$
& $0$ & $\alpha_1$ & $0$ & $\alpha_1$ & $0$ & $\alpha_1$\\
$\ell^{-1}e\cdot(h\times h)$
& $0$ & $\alpha_1$ & $0$ & $\alpha_1$ & $0$ & $\alpha_1$\\
$(3\ell^3)^{-1}e\cdot(e\times e)$
& $0$ & $0$ & $0$ & $0$ & $0$ & $\alpha_1$\\
\midrule
$\CS(\omega)$
& $\alpha_0$ & $\alpha_0$ & $\alpha_0$ & $\alpha_0$ & $\alpha_0$ & $\alpha_0$\\
$2h\cdot R$
& $\alpha_2$ & $\alpha_2$ & $\alpha_2$ & $\alpha_2$ & $\alpha_2$ & $\alpha_2$\\
$h\cdot D_\omega h$
& $0$ & $\alpha_0$ & $\alpha_2$ & $\alpha_2$ & $0$ & $\alpha_2$\\
$\frac13 h\cdot(h\times h)$
& $0$ & $\alpha_2$ & $\alpha_2$ & $\alpha_2$ & $0$ & $\alpha_2$\\
$\ell^{-2}h\cdot(e\times e)$
& $0$ & $0$ & $0$ & $0$ & $0$ & $\alpha_2$\\
$\ell^{-2}e\cdot T$
& $0$ & $0$ & $0$ & $0$ & $\alpha_2$ & $\alpha_2$\\
\end{tabular}
\caption{JPZ Chern--Simons actions.}
\label{tab:lagrangian-content}
\end{table}

The upper block collects the Einstein--Hilbert-like $\alpha_1$ contributions of the standard/gravitational sector, while the lower block contains the $\alpha_0$ and $\alpha_2$ sectors, including the Lorentz Chern--Simons/exotic terms and the additional $h$-dependent contributions.

Two observations are important for what follows. First, all six theories share the same couplings $\alpha_0,\alpha_1,\alpha_2$, but the underlying algebra determines which Lagrangian terms correspond to each coupling. Second, setting one of these couplings to zero restricts the theory to a subspace of coupling space. It does not, by itself, remove the corresponding gauge field or constitute a consistent field truncation. Instead, the rank of the invariant tensor controls the number of independent bulk equations. This distinction provides the organizing principle for the next section.

\section{Bulk dynamics, coupling sectors, and field equations}
\label{sec:bulk-dynamics}

\subsection{Variational equations and the curvature matrix}
\label{subsec:bulk-matrix}

For every algebra, the bulk variation has the universal form
\begin{equation}
\delta I_\sigma=
\frac{k}{2\pi}\int_{\mathcal M}
\langle\delta\mathcal A\wedge\mathcal F\rangle_\sigma
+\delta I_{\partial\mathcal M}\,.
\label{eq:bulk-variation-v3}
\end{equation}
This leads to quite a complicated set of equations that are linear in the curvatures; see Appendix~\ref {app:explicit-actions}. Introducing the curvature vector
\begin{equation}
\mathbf F^a=
\begin{pmatrix}F_J^a\\F_P^a\\F_Z^a\end{pmatrix}\,,
\end{equation}
and suppressing the common factor $\eta_{ab}$ offers a new perspective. In fact, the three Euler--Lagrange equations are
\begin{equation}
\Omega_\sigma\,\mathbf F^a=0\,,
\label{eq:bulk-matrix-v3}
\end{equation}
where the matrices $\Omega_\sigma$ are given in Eq.~\eqref{matrices} in terms of the same three couplings $\alpha_0,\alpha_1,\alpha_2$. Hence the Lie algebra
specifies the curvatures, while the invariant tensor decides which linear
combinations of those curvatures are actually imposed.

We call a coupling sector \emph{regular} when $\Omega_\sigma$ is non-degenerate. More generally, \emph{maximal rank} means the highest rank attainable by $\Omega_\sigma$ for the configuration under consideration. For $B_4$, $\widetilde C_4$, $C_4$, $\mathfrak B_4$, and $\mathfrak C_4$ this rank is three, while for $\widetilde B_4$ it is two.

It is important to distinguish three separate operations. Gauge invariance
of the bulk Chern--Simons action follows from invariance of the bilinear form
and does not require $\Omega_\sigma$ to be non-degenerate; degeneracy only
changes the rank of the variational system. Likewise, restricting the
couplings by imposing special values of one or more $\alpha_i$ preserves the
gauge symmetry, although it may reduce the number of independent field
equations. By contrast, a field truncation such as $h^a=0$ is an additional
dynamical restriction and is consistent only when both the field equations
and the residual gauge transformations preserve it. Thus non-degeneracy is
not a condition for gauge invariance: it is the condition that makes the
variational equations equivalent to curvatures equal to zero.

If $\det\Omega_\sigma\neq0$, the field equations~\eqref{eq:bulk-matrix-v3} are equivalent to
\begin{equation}
F_J^a=F_P^a=F_Z^a=0\,.
\label{eq:full-flatness-v3}
\end{equation}
We refer to this as \emph{full gauge flatness}. It is equivalent to the variational equations only in the regular sectors of coupling space.

The $\Omega_\sigma$ determinants are:
\begin{align}
\det\Omega_{\widetilde B_4}&=0\,,&
\det\Omega_{B_4}&=-2\alpha_1^2(\alpha_0-\alpha_2)\,,\notag\\
\det\Omega_{\widetilde C_4}&=-\alpha_1^2\alpha_2\,,&
\det\Omega_{C_4}&=-\alpha_1^2(\alpha_0-\alpha_2)\,,\notag\\
\det\Omega_{\mathfrak B_4}&=-\alpha_2^3\,,&
\det\Omega_{\mathfrak C_4}&=-(\alpha_0-\alpha_2)(\alpha_1-\alpha_2)(\alpha_1+\alpha_2)\,.
\label{eq:determinants}
\end{align}

The intrinsically singular $\widetilde B_4$, due to its vanishing determinant, has no regular sector. The regular sectors of the coupling space $(\alpha_0,\alpha_1,\alpha_2)$ can be summarized as follows:

\begin{itemize} 
    \item ${B_4}$: $\alpha_1\neq 0$ and $\alpha_0\neq\alpha_2$.
    
    \item ${\widetilde C_4}$: both $\alpha_1\neq 0$ and $\alpha_2\neq 0$.
    
    \item ${C_4}$: $\alpha_1\neq 0$ and $\alpha_0\neq\alpha_2$.
    
    \item ${\mathfrak B_4}$: $\alpha_2\neq 0$.
    
    \item ${\mathfrak C_4}$: $\alpha_0\neq\alpha_2$ and $\alpha_1\neq\pm\alpha_2$.
\end{itemize}

Crossing a determinant zero for any of the other five cases does not change the gauge algebra. It changes only the variational problem by removing one or more independent curvature equations. We collect the lower-rank-two and rank-one configurations separately in Appendix~\ref{app:degenerate}.

\subsection{Field equations for the six JPZ configurations}
\label{subsec:six-field-equations}

Table~\ref{tab:eom-v3} lists the regular/maximal-rank field equations explicitly. 

\begin{table}[H]
\centering
\small
\renewcommand{\arraystretch}{1.35}
\resizebox{\textwidth}{!}{%
\begin{tabular}{c||ccc|l}
Configuration & $J$ equation & $P$ equation & $Z$ equation & coupling condition\\
\midrule
\midrule
$\widetilde B_4$
& $R=0$
& \multicolumn{2}{c|}{$\alpha_1\ell^{-1}T+\alpha_2D_\omega h=0\qquad$}
& $(\alpha_1,\alpha_2)\neq(0,0)$\\
$B_4$
& $R+\frac12h\times h=0$
& $T+h\times e=0$
& $D_\omega h=0$
& $\alpha_1\neq0$, $\alpha_0\neq\alpha_2$\\
$\widetilde C_4$
& $R=0$
& $T=0$
& $D_\omega h+\frac12h\times h=0$
& $\alpha_1\alpha_2\neq0$\\
$C_4$
& $R=0$
& $T+h\times e=0$
& $D_\omega h+\frac12h\times h=0$
& $\alpha_1\neq0$, $\alpha_0\neq\alpha_2$\\
$\mathfrak B_4$
& $R=0$
& $T=0$
& $D_\omega h+\frac1{2\ell^2}e\times e=0$
& $\alpha_2\neq0$\\
$\mathfrak C_4$
& $R=0$
& $T+h\times e=0$
& $D_\omega h+\frac12h\times h+\frac1{2\ell^2}e\times e=0$
& $(\alpha_0-\alpha_2)(\alpha_1^2-\alpha_2^2)\neq0$\\
\end{tabular}%
}
\caption{Independent bulk equations in the regular sectors and in the
maximal-rank $\widetilde B_4$ sector.}
\label{tab:eom-v3}
\end{table}

The table already reveals the main geometrical distinction. The additional
field $h^a$ can source Lorentz curvature, torsion, its own curvature, or the
$e\times e$ sector depending on the commutation rules. The next question
is how to solve these equations
while restoring the original interpretation of $\omega$, $e$, and $h$.

Before solving the field equations, it helps to expose the abstract isomorphisms among several $JPZ$ algebras that greatly simplify the analysis.

\section{Direct-sum isomorphisms}
\label{sec:isomorphisms}

Several resonant $JPZ$ configurations admit direct-sum decompositions after suitable changes of basis \cite{Durka:2016eun,Soroka:2006aj,Soroka:2011tc}. The semisimple algebra $\mathfrak C_4$ provides a well-known example. It is the Soroka--Soroka, or AdS--Lorentz, algebra \cite{Durka:2011nf,Soroka:2006aj,Soroka:2011tc, Salgado:2013eut}, with
\begin{equation}
\mathfrak C_4\simeq\so(2,2)\oplus\so(2,1)\,.
\label{eq:ads-direct-sum-class}
\end{equation}

The central new observation is that the other three apparently different resonant $JPZ$ configurations belong to the same abstract isomorphism class,
\begin{equation}
B_4\simeq\widetilde C_4\simeq C_4
\simeq\iso(2,1)\oplus\so(2,1)\,.
\label{eq:poincare-direct-sum-class}
\end{equation}

We now make these decompositions explicit. Table~\ref{tab:direct-sum-bases} collects the adapted bases and the corresponding gauge connections. All transformations are invertible changes of generators.

\begin{table}[H]
\centering
\small
\renewcommand{\arraystretch}{1.35}
\begin{tabular}{c|c|c|c|c}
Configuration & $L$ & $\mathcal{T}$ & $S$ & adapted connection\\
\midrule
\midrule
$B_4$ & $\frac12(J+Z)$ & $P$ & $\frac12(J-Z)$
& $(\omega+h)L+\ell^{-1}e\,\mathcal{T}+(\omega-h)S$\\
$\widetilde C_4$ & $J-Z$ & $P$ & $Z$
& $\omega L+\ell^{-1}e\,\mathcal{T}+(\omega+h)S$\\
$C_4$ & $Z$ & $P$ & $J-Z$
& $(\omega+h)L+\ell^{-1}e\,\mathcal{T}+\omega S$\\
$\mathfrak C_4$ & $Z$ & $P$ & $J-Z$
& $(\omega+h)L+\ell^{-1}e\,\mathcal{T}+\omega S$
\end{tabular}
\caption{Adapted direct-sum bases for $B_4$, $\widetilde C_4$, $C_4$, and $\mathfrak C_4$.}
\label{tab:direct-sum-bases}
\end{table}
Substituting the first three changes of basis into the original $JPZ$ commutators gives the same algebra in each case:
\begin{equation}
[L,L]\sim L\,,
\qquad
[L,\mathcal{T}]\sim \mathcal{T}\,,
\qquad
[\mathcal{T},\mathcal{T}]=0\,,\label{LT_ISO}
\end{equation}
therefore, for $B_4$, $\widetilde C_4$, and $C_4$, the generators $(L,\mathcal{T})$ span the Poincar\'e sector, 
while $S$ generates a commuting another Lorentz sector,
\begin{equation}
[S,S]\sim S\,.\label{S_Lorentz}
\end{equation}
All mixed brackets involving $S$ vanish. For $\mathfrak C_4$ the only change is
\begin{equation}
[\mathcal{T},\mathcal{T}]\sim L\,,
\end{equation}
so that $(L,\mathcal{T})$ spans the AdS sector.

The corresponding curvatures take a particularly simple form:

\begin{table}[ht]
\centering
\small
\renewcommand{\arraystretch}{1.35}
\begin{tabular}{c||c|c|c}
Configuration & $F_L$ & $\ell F_{\mathcal{T}}$ & $F_S$\\
\midrule
\midrule
$B_4$ & $R(\omega+h)$ & $D_{\omega+h}e$ & $R(\omega-h)$\\
$\widetilde C_4$ & $R(\omega)$ & $D_\omega e$ & $R(\omega+h)$\\
$C_4$ & $R(\omega+h)$ & $D_{\omega+h}e$ & $R(\omega)$\\
$\mathfrak C_4$ & $R(\omega+h)+\frac1{2\ell^2}e\times e$ & $D_{\omega+h}e$ & $R(\omega)$\\
\end{tabular}
\caption{Curvatures in the adapted direct-sum bases.}
\label{tab:direct-sum-curvatures}
\end{table}

We also rewrite the corresponding invariant tensors as:

\begin{table}[ht]
\centering
\small
\renewcommand{\arraystretch}{1.25}
\begin{tabular}{c||c|c|c|c}
Configuration & $\langle L,L\rangle/\eta$ & $\langle L,\mathcal{T}\rangle/\eta$
& $\langle \mathcal{T},\mathcal{T}\rangle/\eta$ & $\langle S,S\rangle/\eta$\\
\midrule
$B_4$ & $\frac12(\alpha_0+\alpha_2)$ & $\alpha_1$ & $0$ &
$\frac12(\alpha_0-\alpha_2)$\\
$\widetilde C_4$ & $\alpha_0-\alpha_2$ & $\alpha_1$ & $0$ & $\alpha_2$\\
$C_4$ & $\alpha_2$ & $\alpha_1$ & $0$ & $\alpha_0-\alpha_2$\\
$\mathfrak C_4$ & $\alpha_2$ & $\alpha_1$ & $\alpha_2$ & $\alpha_0-\alpha_2$
\end{tabular}
\caption{Invariant-tensor entries in the adapted basis.}
\label{tab:direct-sum-couplings}
\end{table}

In the ordered basis $(L,\mathcal{T},S)$, the same change of basis block-diagonalizes the invariant tensors. 
\begin{align}
\Omega^{L\mathcal{T}S}_{B_4}&=
\begin{pmatrix}
\frac12(\alpha_0+\alpha_2)&\alpha_1&0\\
\alpha_1&0&0\\
0&0&\frac12(\alpha_0-\alpha_2)
\end{pmatrix}\,,&
\Omega^{L\mathcal{T}S}_{C_4}&=
\begin{pmatrix}
\alpha_2&\alpha_1&0\\
\alpha_1&0&0\\
0&0&\alpha_0-\alpha_2
\end{pmatrix}\,,\notag\\
\Omega^{L\mathcal{T}S}_{\widetilde C_4}&=
\begin{pmatrix}
\alpha_0-\alpha_2&\alpha_1&0\\
\alpha_1&0&0\\
0&0&\alpha_2
\end{pmatrix}\,,&
\Omega^{L\mathcal{T}S}_{\mathfrak C_4}&=
\begin{pmatrix}
\alpha_2&\alpha_1&0\\
\alpha_1&\alpha_2&0\\
0&0&\alpha_0-\alpha_2
\end{pmatrix}\,.
\end{align}

\subsection{\texorpdfstring{The $C_4$ and $\mathfrak C_4$ split actions}{The C4 and semisimple C4 split actions}}
\label{subsec:C4-Adami-dictionary-v3}

For both $C_4$ and $\mathfrak C_4$, we use
\begin{equation}
\varpi=\omega+h\,,
\qquad
\kappa=\omega\,,\label{eq:split-connections}
\end{equation}
and define
\begin{equation}
\alpha_L=\alpha_2\,,
\qquad
\alpha_{\mathcal{T}}=\alpha_1\,,
\qquad
\alpha_S=\alpha_0-\alpha_2\,.
\end{equation}
The invariant tensors then take the form
\begin{equation}
\Omega_{C_4}^{(L\mathcal{T}S)}
=
\begin{pmatrix}
\alpha_L&\alpha_{\mathcal{T}}&0\\
\alpha_{\mathcal{T}}&0&0\\
0&0&\alpha_S
\end{pmatrix},
\qquad
\Omega_{\mathfrak C_4}^{(L\mathcal{T}S)}
=
\begin{pmatrix}
\alpha_L&\alpha_{\mathcal{T}}&0\\
\alpha_{\mathcal{T}}&\alpha_L&0\\
0&0&\alpha_S
\end{pmatrix}.
\end{equation}
Up to exact forms, the corresponding Lagrangians become
\begin{equation}
\mathcal L_{C_4}
=
\alpha_S\CS(\kappa)
+\alpha_L\CS(\varpi)
+\frac{2\alpha_{\mathcal{T}}}{\ell}e\cdot R(\varpi)\,,
\label{eq:C4-direct-action-v3}
\end{equation}
and
\begin{align}
\mathcal L_{\mathfrak C_4}
={}&
\alpha_S\CS(\kappa)
+\alpha_L\left(
\CS(\varpi)
+\frac1{\ell^2}e\cdot D_\varpi e
\right)\nonumber\\
&+\frac{2\alpha_{\mathcal{T}}}{\ell}
\left(
e\cdot R(\varpi)
+\frac1{6\ell^2}e\cdot(e\times e)
\right)\,.
\label{eq:mathfrakC4-direct-action}
\end{align}

These shifted variables will also simplify the integration of the field equations in the next section. They do not redefine the original gravitational embedding: for us $\omega$ should always remain the true Lorentz spin connection associated with the original Lorentz $J_a$ generator. In sectors where $e^a$ is invertible and $D_\varpi e=0$, the shifted connection $\varpi=\omega+h$ is the Levi--Civita connection of the metric defined by $e^a$.

Interestingly, within $\mathfrak C_4$, the AdS sector can be decomposed further, exposing the three commuting Lorentz sectors.

\subsection{\texorpdfstring{Chiral decomposition of $\mathfrak C_4$}{Chiral decomposition of mathfrak C4}}
\label{subsec:chiral-decomposition}

As discussed above, the generators $(L_a,\mathcal{T}_a)$ span the AdS sector, while $S_a$ generates the commuting Lorentz sector,
\begin{equation}
\mathfrak C_4
\simeq
\so(2,2)_{L\mathcal{T}}\oplus\so(2,1)_S\,.
\end{equation}
Using the standard left/right, or chiral, decomposition of the three-dimensional AdS algebra \cite{Achucarro:1986uwr,Witten:1988hc,Concha:2018jjj}, we define
\begin{equation}
J_a^\pm=\frac12\left(L_a\pm \mathcal{T}_a\right),
\qquad
J_a^S=S_a\,.
\label{eq:mathfrakC4-chiral-generators}
\end{equation}
The commutation relations then take the form
\begin{equation}
[J_a^+,J_b^+]
=
\epsilon_{ab}{}^cJ_c^+\,,
\qquad
[J_a^-,J_b^-]
=
\epsilon_{ab}{}^cJ_c^-\,,
\qquad
[J_a^S,J_b^S]
=
\epsilon_{ab}{}^cJ_c^S\,,
\end{equation}
with
\begin{equation}
[J_a^+,J_b^-]
=
[J_a^\pm,J_b^S]
=
0\,.
\label{eq:mathfrakC4-chiral-algebra}
\end{equation}
Hence the three sectors commute independently and
\begin{equation}
\mathfrak C_4
\simeq
\so(2,1)_+\oplus\so(2,1)_-\oplus\so(2,1)_S\,.
\label{eq:mathfrakC4-triple-direct-sum}
\end{equation}
The same decomposition is visible directly at the level of the gauge connection,
\begin{equation}
\mathcal A
=
(\omega+h)^aL_a
+\frac1\ell e^aT_a
+\omega^aS_a =
A^{+a}J_a^+
+
A^{-a}J_a^-
+
A^{Sa}J_a^S\,,
\label{eq:mathfrakC4-triple-connection}
\end{equation}
where
\begin{equation}
A^\pm
=
(\omega+h)\pm\frac1\ell e\,,
\qquad
A^S=\omega\,.
\label{eq:triple-lorentz-connections}
\end{equation}
Its curvature therefore separates into three independent Lorentz curvatures,
\begin{equation}
\mathcal F
=
R(A^+)^aJ_a^+
+
R(A^-)^aJ_a^-
+
R(A^S)^aJ_a^S\,.
\label{eq:mathfrakC4-triple-curvature}
\end{equation}
In the regular sector, the bulk equations are simply
\begin{equation}
R(A^+)=0\,,
\qquad
R(A^-)=0\,,
\qquad
R(A^S)=0\,.
\end{equation}
The invariant tensor diagonalizes in the same basis,
\begin{equation}
\langle J^+,J^+\rangle
=
\frac12(\alpha_2+\alpha_1)\eta\,,
\qquad
\langle J^-,J^-\rangle
=
\frac12(\alpha_2-\alpha_1)\eta\,,
\qquad
\langle J^S,J^S\rangle
=
(\alpha_0-\alpha_2)\eta\,,
\label{eq:mathfrakC4-chiral-levels}
\end{equation}
with all mixed pairings vanishing. Accordingly the
$\mathfrak C_4$ Chern--Simons Lagrangian splits into three independent terms,
\begin{equation}
\mathcal L_{\mathfrak C_4}
=
\frac{\alpha_2+\alpha_1}{2}\CS(A^+)
+
\frac{\alpha_2-\alpha_1}{2}\CS(A^-)
+
(\alpha_0-\alpha_2)\CS(A^S)\,.
\label{eq:mathfrakC4-triple-action}
\end{equation}

Thus the decomposition in Eq.~\eqref{eq:mathfrakC4-triple-direct-sum}
is realized simultaneously at the level of the generators, gauge connection,
curvature, invariant tensor, and Chern--Simons action.

\subsection{\texorpdfstring{$JP$ versus $JPZ$}{JP versus JPZ}}

The direct-sum structure of the $JPZ$ algebras can be compared with the
standard two-generator case. For AdS$_3$, the familiar decomposition
\begin{equation}
\so(2,2)
\simeq
\so(2,1)_+\oplus\so(2,1)_-\,,
\end{equation}
underlies the Chern--Simons formulation of gravity
\cite{Achucarro:1986uwr,Witten:1988hc}, with the corresponding chiral
connections
\begin{equation}
A^\pm=\omega\pm\frac1\ell e\,.
\end{equation}
By contrast, the Poincar\'e algebra $\iso(2,1)$ is non-semisimple and does
not decompose.

The $JPZ$ extensions introduce a third generator sector. In the semisimple
case $\mathfrak C_4$, the generators $(L,\mathcal{T})$ reproduce the AdS algebra,
while $S$ provides an additional commuting Lorentz sector. The six resonant
$JPZ$ configurations can then be organized as
\begin{itemize}
    \item \(\widetilde B_4\)
    \quad (intrinsically degenerate invariant tensor);

    \item \(\mathfrak B_4\)
    \quad (Maxwell algebra);

    \item \(\mathfrak C_4
    \simeq\so(2,2)_{L\mathcal{T}}\oplus\so(2,1)_S
    \simeq\so(2,1)_+\oplus\so(2,1)_-\oplus\so(2,1)_S\)
    \quad (AdS $\oplus$ Lorentz);

    \item \(B_4\simeq\widetilde C_4\simeq C_4
    \simeq\iso(2,1)_{L\mathcal{T}}\oplus\so(2,1)_S\)
    \quad (Poincar\'e $\oplus$ Lorentz).
\end{itemize}
The adapted bases expose the abstract direct-sum structure, but they do
not replace the original gravitational embedding. In the following section
we use them only to integrate the field equations and then translate the
result back to the original variables $(\omega,e,h)$.

A systematic treatment of more general generator redefinitions is left for future work. Such a framework should reconcile the direct-sum structures above, teleparallel and Mielke--Baekler realizations, and the usual dS/AdS/Poincar\'e branches, while also clarifying the number of apparently distinct resonant configurations associated with invertible changes of generators. 

\section{Solving the six JPZ systems}
\label{sec:local-solutions-v3}

\subsection{Integration strategy and the original gravitational embedding}
\label{subsec:integration-strategy-v3}

We always expand the gauge connection in the original $JPZ$ basis of
Eq.~\eqref{eq:connection_def}. The direct-sum variables used below serve only
to integrate the field equations; in each case we return to the original
fields $(\omega,e,h)$.

The following two local observations will be used repeatedly in solving the
six systems. First, the flatness condition
\begin{equation}
R(\Gamma)=0
\label{eq:pure-gauge-v3}
\end{equation}
implies that the Lorentz connection is just pure gauge,
\begin{equation}
\Gamma=U^{-1}dU\,,
\end{equation}
where $U$ is a local Lorentz-group element. If, in addition,
$D_\Gamma e=0$, then $d(Ue)=0$. Therefore, locally, there exist functions
$X^a$ such that $(Ue)^a=dX^a$, or equivalently
\begin{equation}
e^a=(U^{-1})^a{}_b\,dX^b\,.
\label{eq:cov-constant-e-v3}
\end{equation}
The $X^a$ are local functions; when the dreibein is invertible, they
may be used as local Minkowski coordinates. In the gauge $\Gamma=0$, this
reduces to $e^a=dX^a$.

Second, the shifted connection $\varpi=\omega+h$ defined in
Eq.~\eqref{eq:split-connections} will repeatedly turn the original
$JPZ$ equations into standard curvature and torsion conditions. In
particular,
\begin{align}
R(\varpi)&=R(\omega)+D_\omega h+\frac12h\times h\,,
\label{eq:shift-R-v3}\\
D_{\varpi}e&=T(\omega,e)+h\times e\,.
\label{eq:shift-T-v3}
\end{align}

\subsection{\texorpdfstring{$\widetilde B_4$}: one determined combination}

In the maximal-rank sector,
$\operatorname{rank}\Omega_{\widetilde B_4}=2$, and we have
\begin{equation}
R(\omega)=0\,,
\qquad
D_\omega\left(\frac{\alpha_1}{\ell}e+\alpha_2h\right)=0\,.
\end{equation}
For $\alpha_2\neq0$, the local solution is
\begin{equation}
\omega=U^{-1}dU\,,
\qquad
e=e_{\rm free}\,,
\qquad
h=\frac1{\alpha_2}
\left(U^{-1}dY-\frac{\alpha_1}{\ell}e_{\rm free}\right)\,.
\label{eq:tildeB-solution-v3}
\end{equation}
The arbitrary invertible one-form $e_{\rm free}$ reflects the intrinsic rank deficiency of the variational system.

For the remaining maximal-rank branch,
$\alpha_2=0$ and $\alpha_1\neq0$,
\begin{equation}
R(\omega)=0\,,
\qquad
T=0\,,
\end{equation}
and hence
\begin{equation}
\omega=U^{-1}dU\,,
\qquad
e=U^{-1}dX\,,
\end{equation}
while $h$ remains unconstrained by the bulk equations.

\subsection{\texorpdfstring{$B_4$}: two flat Lorentz connections}

After introducing
\begin{equation}
\varpi_\pm:=\omega\pm h\,,
\end{equation}
the field equations become
\begin{equation}
R(\varpi_+)=0\,,
\qquad
D_{\varpi_+}e=0\,,
\qquad
R(\varpi_-)=0\,,
\end{equation}
with local solution
\begin{equation}
\varpi_+=U_+^{-1}dU_+\,,
\qquad
e=U_+^{-1}dX\,,
\qquad
\varpi_-=U_-^{-1}dU_-\,.
\end{equation}
In the original variables we end up with
\begin{equation}
\omega=\frac12\left(U_+^{-1}dU_+ +U_-^{-1}dU_-\right)\,,
\qquad
e=U_+^{-1}dX\,,
\qquad
h=\frac12\left(U_+^{-1}dU_+ -U_-^{-1}dU_-\right)\,.
\label{eq:B4-solution-v3}
\end{equation}
The solution is locally Minkowski, with $\varpi_+$ as its torsionless connection, although $R(\omega)$ need not vanish.

\subsection{\texorpdfstring{$\widetilde C_4$}: the original Poincar\'e pair}

The equations can be written as
\begin{equation}
R(\omega)=0\,,
\qquad
D_\omega e=0\,,
\qquad
R(\varpi)=0\,,
\end{equation}
and therefore
\begin{equation}
\omega=U_0^{-1}dU_0\,,
\qquad
e=U_0^{-1}dX\,,
\qquad
\varpi=U_1^{-1}dU_1\,,
\end{equation}
with
\begin{equation}
h=U_1^{-1}dU_1-U_0^{-1}dU_0\,.
\label{eq:tildeC-solution-v3}
\end{equation}
Thus the original $\omega$ is the torsionless flat connection of the dreibein $e$.

\subsection{\texorpdfstring{$C_4$}: Poincar\'e gravity plus Lorentz}

Here
\begin{equation}
R(\varpi)=0\,,
\qquad
D_{\varpi}e=0\,,
\qquad
R(\omega)=0\,,\label{eq:C4-direct-eom}
\end{equation}
so locally we write
\begin{equation}
\omega=U_S^{-1}dU_S\,,
\qquad
e=U_L^{-1}dX\,,
\qquad
\varpi=U_L^{-1}dU_L\,,
\end{equation}
and
\begin{equation}
h=U_L^{-1}dU_L-U_S^{-1}dU_S\,.
\label{eq:C4-solution-v3}
\end{equation}
The solution is locally flat, with Levi--Civita connection
$\varpi=\omega+h$. We return to this case in a further section.

\subsection{\texorpdfstring{$\mathfrak B_4$: Maxwell field}{B4: Maxwell field}}

The equations are
\begin{equation}
R(\omega)=0\,,
\qquad
T=0\,,
\qquad
D_\omega h+\frac1{2\ell^2}e\times e=0\,.
\end{equation}
The first two define a locally flat torsionless Poincar\'e pair,
\begin{equation}
\omega=U^{-1}dU\,,
\qquad
e=U^{-1}dX\,.
\end{equation}
To integrate the remaining equation, define the Lorentz-rotated one-form
\begin{equation}
\widetilde h:=Uh\,.
\end{equation}
Using $\omega=U^{-1}dU$ and $e=U^{-1}dX$, Lorentz covariance gives
\begin{equation}
D_\omega h=U^{-1}d\widetilde h\,,
\qquad
e\times e=U^{-1}(dX\times dX)\,.
\end{equation}
Hence the last field equation reduces to
\begin{equation}
d\widetilde h
=
-\frac1{2\ell^2}dX\times dX\,.
\end{equation}
Since
\begin{equation}
d(X\times dX)=dX\times dX\,,
\end{equation}
the local solution is
\begin{equation}
\widetilde h
=
dY-\frac1{2\ell^2}X\times dX\,,
\end{equation}
where $Y^a$ are local functions. Therefore, after reverting to the original variables,
\begin{equation}
\omega=U^{-1}dU\,,
\qquad
e=U^{-1}dX\,,
\qquad
h=U^{-1}\left(
dY-\frac1{2\ell^2}X\times dX
\right)\,.
\label{eq:mathfrakB-solution-v3}
\end{equation}
Thus the metric is locally Minkowski, while the one-form $h$ need not vanish
in the original $JPZ$ decomposition.

\subsection{\texorpdfstring{$\mathfrak C_4$: three Lorentz sectors}{mathfrak C4: three Lorentz sectors}}

The field equations are
\begin{equation}
R(\omega)=0\,,
\qquad
D_{\varpi}e=0\,,
\qquad
R(\varpi)+\frac1{2\ell^2}e\times e=0\,.\label{eq:mathfrakC4-local-eom}
\end{equation}
Using the chiral connections $A^\pm$ defined in
Eq.~\eqref{eq:triple-lorentz-connections}, the last two equations become
\begin{equation}
R(A^+)=0\,,
\qquad
R(A^-)=0\,.
\end{equation}
Hence
\begin{equation}
\omega=U_0^{-1}dU_0\,,
\qquad
A^\pm=U_\pm^{-1}dU_\pm\,,
\end{equation}
and therefore
\begin{equation}
\varpi=
\frac12\left(U_+^{-1}dU_+ +U_-^{-1}dU_-\right)\,,
\qquad
e=
\frac\ell2\left(U_+^{-1}dU_+-U_-^{-1}dU_-\right)\,.
\end{equation}
Remembering that $h=\varpi-\omega$, the original fields are
\begin{align}
\omega&=U_0^{-1}dU_0\,,\notag\\
e&=\frac\ell2\left(U_+^{-1}dU_+-U_-^{-1}dU_-\right)\,,\notag\\
h&=\frac12\left(U_+^{-1}dU_+ +U_-^{-1}dU_-\right)-U_0^{-1}dU_0\,.
\label{eq:mathfrakC-solution-v3}
\end{align}
For an invertible dreibein, $\varpi$ is the Levi--Civita connection, and
the last equation in Eq.~\eqref{eq:mathfrakC4-local-eom} shows that the
metric is locally AdS$_3$. Equivalently, the chiral connections
$A^\pm=\varpi\pm\ell^{-1}e$ are flat.

\section{From local geometry to global solutions}
\label{sec:standard-geometries-v3}

The solutions of the previous section characterize the local metric geometry
of the six resonant $JPZ$ algebras. Local flatness or constant curvature,
however, does not fix the global spacetime. Different identifications,
holonomies, and boundary conditions may lead to inequivalent solutions with
the same local geometry. The BTZ black hole is the familiar example: its
metric is locally AdS$_3$, while its black-hole character arises from a
global identification.

In the original $JPZ$ gravitational embedding, the local results are:
\begin{itemize}
    \item For $\widetilde B_4$, the metric is undetermined in the maximal-rank
    branch $\alpha_2\neq0$, whereas $\alpha_2=0$, $\alpha_1\neq0$ gives a
    locally Minkowski metric.
    \item $B_4$, $\widetilde C_4$, $C_4$, and $\mathfrak B_4$ give locally
    Minkowski solution.
    \item $\mathfrak C_4$ gives a locally AdS$_3$.
\end{itemize}
Our aim is not to repeat the asymptotic analyses, but to relate these local
results to known global solutions and boundary symmetries.

The Maxwell theory associated with $\mathfrak B_4$ was considered in
Ref.~\cite{Hoseinzadeh:2014bla}, while its boundary conditions, conserved
charges, and asymptotic symmetries were studied in
Ref.~\cite{Concha:2018zeb}. The resulting algebra is a Maxwell extension of
$\widehat{\mathfrak{bms}}_3$. More recent developments include spin-3 Maxwell
asymptotics \cite{Concha:2024rac} and an explicit description of the
induced boundary dynamics \cite{Hofenstock:2025xoc}.

For the semisimple Soroka--Soroka algebra $\mathfrak C_4$, AdS- and BTZ-type
solutions were discussed in Ref.~\cite{Hoseinzadeh:2014bla}. A BMS-like
ansatz encompassing the relevant stationary solutions, together with
consistent boundary conditions, was developed in
Ref.~\cite{Concha:2018jjj}. Its asymptotic symmetry is a semisimple
enlargement of $\mathfrak{bms}_3$, isomorphic to $\mathfrak{vir}\oplus\mathfrak{vir}\oplus\mathfrak{vir}$. This agrees with the direct-sum decomposition in
Eq.~\eqref{eq:ads-direct-sum-class}. The flat limit was also examined at
the level of the action, field equations, solutions, and asymptotic algebra.

The remaining isomorphic class,
$B_4\simeq\widetilde C_4\simeq C_4$, provides a connection with the Maxwell
Chern--Simons theory with torsion of Ref.~\cite{Adami:2020xkm}. This theory
uses a non-standard basis of the same
$\mathfrak{iso}(2,1)\oplus\mathfrak{so}(2,1)$ algebra appearing in
Eq.~\eqref{eq:poincare-direct-sum-class}. Its action admits a
Mielke--Baekler sector, while BMS-like boundary conditions yield
$\widehat{\mathfrak{bms}}_3\oplus\mathfrak{vir}$ with three independent
central charges. Global solutions, including torsional BTZ-like black holes,
and their conserved charges and thermodynamics were subsequently analyzed
in Ref.~\cite{Aviles:2025nah}. Related non-relativistic torsional
supergravity constructions have also been developed within the
Mielke--Baekler framework \cite{Barriga:2026awj}.

These results provide useful global representatives for the
Poincar\'e+Lorentz class, but the metric interpretation depends on the
choice of dreibein. In particular, the torsional realization uses
$\bar e^a=-\lambda h^a$, rather than the original $JPZ$ dreibein $e^a$.
Thus its BTZ-like metric need not coincide with the locally flat metric
constructed from $e^a$ in the regular $C_4$ sector. The algebraic
isomorphism relates the gauge descriptions, not automatically their
boundary charges or gravitational interpretations.

\section{Conclusions and outlook}
\label{sec:discussion-v3}

We have constructed three-dimensional Chern--Simons gravity for the six resonant $JPZ$ algebras, keeping the original identification of the Lorentz connection $\omega$, dreibein $e$, and additional field $h$ throughout. Their commutators determine the curvature components, while the invariant
tensors select the combinations imposed by the bulk field equations. The matrix form $\Omega_\sigma\mathbf F=0$ makes this distinction transparent and identifies the maximal-rank coupling sectors.

Our main algebraic result is the threefold isomorphism
$B_4\simeq\widetilde C_4\simeq C_4\simeq \iso(2,1)\oplus\so(2,1)$. The corresponding changes of
basis simplify the field equations, but do not preserve the original gravitational embedding and interpretation of what is the Lorentz generator and spin connection. 

In the non-degenerate sectors, bulk equations reduce to full gauge flatness, but the geometrical content of this condition depends on how $h^a$ enters the curvatures. The $B_4$, $\widetilde C_4$, $C_4$, and $\mathfrak B_4$ sectors give locally flat metrics in the original embedding, whereas $\mathfrak C_4$ gives locally AdS$_3$ geometry. The configuration $\widetilde B_4$ is exceptional because its invariant tensor is intrinsically rank deficient: for the generic maximal-rank branch with $\alpha_2\neq0$, the metric remains underdetermined, while the special branch $\alpha_2=0$, $\alpha_1\neq0$ is locally flat. These local metric classes do not determine the global solutions by themselves; non-trivial holonomies, identifications, and boundary conditions provide the additional global data.

A natural continuation is to explore more general invertible redefinitions of the $JPZ$ generators together with their invariant tensors and gravitational embeddings, also in resonant superalgebras. Such a classification should clarify the relation among the direct-sum, teleparallel and Mielke--Baekler descriptions, and determine which resonant configurations or gravitational realizations remain genuinely distinct once the full Chern--Simons data are taken into account.

\section{Acknowledgments}

R.D. thanks the Instituto de Ciencias Exactas y Naturales (ICEN), Universidad Arturo Prat, for its hospitality during his visit. The work of R.D. was supported by the program ``IDUB Excellence Initiative--Research University'' for the years 2020--2026 at the University of Wroc\l{}aw. N.M. was funded by the National Agency for Research and Development (ANID) through FONDECYT Regular Grant No.~1231133. D.M.P. acknowledges financial support from the Chilean government through FONDECYT Grant No.~11240533.

\appendix
\section{Explicit JPZ Chern–Simons Lagrangians and variational equations}
\label{app:explicit-actions}

For convenience, we recall all the  $JPZ$ commutation relations.

\begin{table}[H]
\centering
\renewcommand{\arraystretch}{1.22}
\begin{minipage}[t]{0.5\textwidth}
\centering
\begin{tabular}{c|ccc}
$\widetilde B_4$ & $J$ & $P$ & $Z$\\ \hline
$J$ & $J$ & $P$ & $Z$\\
$P$ & $P$ & $0$ & $0$\\
$Z$ & $Z$ & $0$ & $0$
\end{tabular}
\end{minipage}\hfill
\begin{minipage}[t]{0.5\textwidth}
\centering
\begin{tabular}{c|ccc}
$B_4$ & $J$ & $P$ & $Z$\\ \hline
$J$ & $J$ & $P$ & $Z$\\
$P$ & $P$ & $0$ & $P$\\
$Z$ & $Z$ & $P$ & $J$
\end{tabular}
\end{minipage}

\vspace{5mm}
\begin{minipage}[t]{0.5\textwidth}
\centering
\begin{tabular}{c|ccc}
$\widetilde C_4$ & $J$ & $P$ & $Z$\\ \hline
$J$ & $J$ & $P$ & $Z$\\
$P$ & $P$ & $0$ & $0$\\
$Z$ & $Z$ & $0$ & $Z$
\end{tabular}
\end{minipage}\hfill
\begin{minipage}[t]{0.5\textwidth}
\centering
\begin{tabular}{c|ccc}
$C_4$ & $J$ & $P$ & $Z$\\ \hline
$J$ & $J$ & $P$ & $Z$\\
$P$ & $P$ & $0$ & $P$\\
$Z$ & $Z$ & $P$ & $Z$
\end{tabular}
\end{minipage}

\vspace{5mm}
\begin{minipage}[t]{0.5\textwidth}
\centering
\begin{tabular}{c|ccc}
$\mathfrak B_4$ & $J$ & $P$ & $Z$\\ \hline
$J$ & $J$ & $P$ & $Z$\\
$P$ & $P$ & $Z$ & $0$\\
$Z$ & $Z$ & $0$ & $0$
\end{tabular}
\end{minipage}\hfill
\begin{minipage}[t]{0.5\textwidth}
\centering
\begin{tabular}{c|ccc}
$\mathfrak C_4$ & $J$ & $P$ & $Z$\\ \hline
$J$ & $J$ & $P$ & $Z$\\
$P$ & $P$ & $Z$ & $P$\\
$Z$ & $Z$ & $P$ & $Z$
\end{tabular}
\end{minipage}
\caption{Commutation tables for the six $JPZ$ configurations.}
\label{tab:appendix-jpz-commutators}
\end{table}

The six Chern--Simons Lagrangians below use the action definition of Eq.~\eqref{eq:cs-action}. After each Lagrangian we display the three bulk variational equations that coincide with the three rows of
$\Omega_\sigma(F_J,F_P,F_Z)^T=0$.

\begin{itemize}
\item For $\widetilde B_4$ the Lagrangian is:
\end{itemize}
\begin{equation}
\mathcal L_{\widetilde B_4}
=
\alpha_0\CS(\omega)
+\frac{2\alpha_1}{\ell}e_a\wedge R^a
+2\alpha_2h_a\wedge R^a\,.
\end{equation}
The three variational equations are
\begin{equation}
\left\{
\begin{aligned}
0&=
\alpha_0R^a
+\frac{\alpha_1}{\ell}T^a
+\alpha_2D_\omega h^a\,,
\\
0&=
\alpha_1R^a\,,
\\
0&=
\alpha_2R^a\,.
\end{aligned}
\right.
\end{equation}
For $(\alpha_1,\alpha_2)\neq(0,0)$ these reduce to the maximal-rank system
given in Table~\ref{tab:eom-v3}.

\begin{itemize}
\item For $B_4$, the Lagrangian is:
\end{itemize}
\begin{align}
\mathcal L_{B_4}
={}&
\alpha_0\left(
\CS(\omega)+h_a\wedge D_\omega h^a
\right)
\nonumber\\
&+\alpha_1\left(
\frac{2}{\ell}e_a\wedge R^a
+\frac{2}{\ell}e_a\wedge D_\omega h^a
+\frac1\ell e_a\wedge(h\times h)^a
\right)
\nonumber\\
&+\alpha_2\left(
2h_a\wedge R^a
+\frac13h_a\wedge(h\times h)^a
\right)\,.
\end{align}
The three variational equations are
\begin{equation}
\left\{
\begin{aligned}
0={}&
\alpha_0\left(R^a+\frac12(h\times h)^a\right)
+\frac{\alpha_1}{\ell}\left(T^a+(h\times e)^a\right)
+\alpha_2D_\omega h^a\,,
\\
0={}&
\alpha_1\left(
R^a+\frac12(h\times h)^a+D_\omega h^a
\right)\,,
\\
0={}&
\alpha_2\left(R^a+\frac12(h\times h)^a\right)
+\frac{\alpha_1}{\ell}\left(T^a+(h\times e)^a\right)
+\alpha_0D_\omega h^a\,.
\end{aligned}
\right.
\end{equation}

\begin{itemize}
\item For $\widetilde C_4$ the Lagrangian is:
\end{itemize}
\begin{align}
\mathcal L_{\widetilde C_4}
={}&
\alpha_0\CS(\omega)
+\frac{2\alpha_1}{\ell}e_a\wedge R^a
\nonumber\\
&+\alpha_2\left(
2h_a\wedge R^a
+h_a\wedge D_\omega h^a
+\frac13h_a\wedge(h\times h)^a
\right)\,.
\end{align}
The three variational equations are
\begin{equation}
\left\{
\begin{aligned}
0={}&
\alpha_0R^a
+\frac{\alpha_1}{\ell}T^a
+\alpha_2\left(
R^a+D_\omega h^a+\frac12(h\times h)^a
\right)\,,
\\
0={}&
\alpha_1R^a\,,
\\
0={}&
\alpha_2\left(
R^a+D_\omega h^a+\frac12(h\times h)^a
\right)\,.
\end{aligned}
\right.
\end{equation}

\begin{itemize}
\item For $C_4$ the Lagrangian is:
\end{itemize}
\begin{align}
\mathcal L_{C_4}
={}&
\alpha_0\CS(\omega)
\nonumber\\
&+\alpha_1\left(
\frac{2}{\ell}e_a\wedge R^a
+\frac{2}{\ell}e_a\wedge D_\omega h^a
+\frac1\ell e_a\wedge(h\times h)^a
\right)
\nonumber\\
&+\alpha_2\left(
2h_a\wedge R^a
+h_a\wedge D_\omega h^a
+\frac13h_a\wedge(h\times h)^a
\right)\,.
\end{align}
The three variational equations are
\begin{equation}
\left\{
\begin{aligned}
0={}&
\alpha_0R^a
+\frac{\alpha_1}{\ell}\left(T^a+(h\times e)^a\right)
+\alpha_2\left(
D_\omega h^a+\frac12(h\times h)^a
\right)\,,
\\
0={}&
\alpha_1\left(
R^a+D_\omega h^a+\frac12(h\times h)^a
\right)\,,
\\
0={}&
\alpha_2R^a
+\frac{\alpha_1}{\ell}\left(T^a+(h\times e)^a\right)
+\alpha_2\left(
D_\omega h^a+\frac12(h\times h)^a
\right)\,.
\end{aligned}
\right.
\end{equation}

\begin{itemize}
\item For $\mathfrak B_4$ the Lagrangian is:
\end{itemize}
\begin{equation}
\mathcal L_{\mathfrak B_4}
=
\alpha_0\CS(\omega)
+\frac{2\alpha_1}{\ell}e_a\wedge R^a
+\alpha_2\left(
2h_a\wedge R^a+\frac1{\ell^2}e_a\wedge T^a
\right)\,.
\end{equation}
The three variational equations are
\begin{equation}
\left\{
\begin{aligned}
0={}&
\alpha_0R^a
+\frac{\alpha_1}{\ell}T^a
+\alpha_2\left(
D_\omega h^a+\frac1{2\ell^2}(e\times e)^a
\right)\,,
\\
0={}&
\alpha_1R^a+\frac{\alpha_2}{\ell}T^a\,,
\\
0={}&
\alpha_2R^a\,.
\end{aligned}
\right.
\end{equation}

\begin{itemize}
\item For $\mathfrak C_4$ the Lagrangian is:
\end{itemize}
\begin{align}
\mathcal L_{\mathfrak C_4}
={}&
\alpha_0\CS(\omega)
\nonumber\\
&+\alpha_1\left(
\frac{2}{\ell}e_a\wedge R^a
+\frac{2}{\ell}e_a\wedge D_\omega h^a
+\frac1{3\ell^3}e_a\wedge(e\times e)^a
+\frac1\ell e_a\wedge(h\times h)^a
\right)
\nonumber\\
&+\alpha_2\left(
2h_a\wedge R^a
+\frac1{\ell^2}e_a\wedge T^a
+h_a\wedge D_\omega h^a
+\frac1{\ell^2}h_a\wedge(e\times e)^a
+\frac13h_a\wedge(h\times h)^a
\right)\,.
\end{align}
The three variational equations are
\begin{equation}
\left\{
\begin{aligned}
0={}&
\alpha_0R^a
+\frac{\alpha_1}{\ell}\left(T^a+(h\times e)^a\right)
+\alpha_2\left(
D_\omega h^a+\frac12(h\times h)^a
+\frac1{2\ell^2}(e\times e)^a
\right)\,,
\\
0={}&
\alpha_1R^a
+\frac{\alpha_2}{\ell}\left(T^a+(h\times e)^a\right)
+\alpha_1\left(
D_\omega h^a+\frac12(h\times h)^a
+\frac1{2\ell^2}(e\times e)^a
\right)\,,
\\
0={}&
\alpha_2R^a
+\frac{\alpha_1}{\ell}\left(T^a+(h\times e)^a\right)
+\alpha_2\left(
D_\omega h^a+\frac12(h\times h)^a
+\frac1{2\ell^2}(e\times e)^a
\right)\,.
\end{aligned}
\right.
\end{equation}

For every non-degenerate invariant tensor, these three equations are equivalent
to the vanishing of the three gauge-curvature components. In the degenerate coupling sectors, some of the three equations become linearly dependent. Such
cases are classified in Appendix~\ref{app:degenerate}.

\section{Degenerate coupling configurations}
\label{app:degenerate}

When $\det\Omega_\sigma=0$, the gauge algebra and the Chern--Simons construction remain unchanged, but the invariant tensor has lower rank and the action imposes fewer independent curvature equations. A rank-two
invariant tensor gives two independent Lorentz-vector equations, while a rank-one invariant tensor leaves only one. At $\alpha_0=\alpha_1=\alpha_2=0\,,$ the invariant tensor has rank zero and the bulk action imposes no curvature equation.

The tables below collect all non-zero degenerate coupling sectors. In the rank-two table, the stated
conditions exclude the lower-rank intersections.

\subsection*{Rank-two sectors}

\begin{table}[H]
\centering
\small
\renewcommand{\arraystretch}{1.25}
\begin{tabular}{c||c|l}
Configuration & rank-two condition & independent equations\\
\midrule
\midrule
$\widetilde B_4$
& $(\alpha_1,\alpha_2)\neq(0,0)$
& $F_J=0$, $\alpha_1F_P+\alpha_2F_Z=0$\\
$B_4$
& $\alpha_1\neq0$, $\alpha_0=\alpha_2$
& $F_P=0$, $F_J+F_Z=0$\\
$B_4$
& $\alpha_1=0$, $\alpha_0^2\neq\alpha_2^2$
& $F_J=0$, $F_Z=0$\\
$\widetilde C_4$
& $\alpha_1\neq0$, $\alpha_2=0$
& $F_J=0$, $F_P=0$\\
$\widetilde C_4$
& $\alpha_1=0$, $\alpha_2\neq0$, $\alpha_0\neq\alpha_2$
& $F_J=0$, $F_Z=0$\\
$C_4$
& $\alpha_1\neq0$, $\alpha_0=\alpha_2$
& $F_P=0$, $F_J+F_Z=0$\\
$C_4$
& $\alpha_1=0$, $\alpha_2\neq0$, $\alpha_0\neq\alpha_2$
& $F_J=0$, $F_Z=0$\\
$\mathfrak B_4$
& $\alpha_2=0$, $\alpha_1\neq0$
& $F_J=0$, $F_P=0$\\
$\mathfrak C_4$
& $\alpha_0=\alpha_2$, $\alpha_1^2\neq\alpha_2^2$
& $F_P=0$, $F_J+F_Z=0$\\
$\mathfrak C_4$
& $\alpha_1=\alpha_2\neq0$, $\alpha_0\neq\alpha_2$
& $F_J=0$, $F_P+F_Z=0$\\
$\mathfrak C_4$
& $\alpha_1=-\alpha_2\neq0$, $\alpha_0\neq\alpha_2$
& $F_J=0$, $F_P-F_Z=0$\\
\end{tabular}
\caption{Rank-two degenerate coupling sectors.}
\label{tab:rank-two-v3}
\end{table}

\subsection*{Rank-one sectors}

At rank one, the action imposes a single curvature equation:
\begin{table}[H]
\centering
\small
\renewcommand{\arraystretch}{1.25}
\begin{tabular}{c||c|l}
Configuration & rank-one condition & independent equation\\
\midrule
\midrule
$\widetilde B_4,\widetilde C_4,C_4,\mathfrak B_4,\mathfrak C_4$
& $\alpha_1=\alpha_2=0$, $\alpha_0\neq0$
& $F_J=0$\\
$B_4$
& $\alpha_1=0$, $\alpha_0=\alpha_2\neq0$
& $F_J+F_Z=0$\\
$B_4$
& $\alpha_1=0$, $\alpha_0=-\alpha_2\neq0$
& $F_J-F_Z=0$\\
$\widetilde C_4,C_4$
& $\alpha_1=0$, $\alpha_0=\alpha_2\neq0$
& $F_J+F_Z=0$\\
$\mathfrak C_4$
& $\alpha_0=\alpha_1=\alpha_2\neq0$
& $F_J+F_P+F_Z=0$\\
$\mathfrak C_4$
& $\alpha_0=\alpha_2=-\alpha_1\neq0$
& $F_J-F_P+F_Z=0$\\
\end{tabular}
\caption{Rank-one degenerate coupling sectors.}
\label{tab:rank-one-v3}
\end{table}

These degenerate sectors are properties of the invariant bilinear form, not
of the underlying Lie algebra. Moving to one of these coupling choices does
not remove any gauge field. This only reduces the number of independent
curvature equations imposed by the action. Setting a field such as $h$ to
zero is a separate restriction.

\section{Poincar\'e, (A)dS, and Mielke--Baekler gravity}
\label{app:cs-poincare_ads}

This appendix, while focusing on $JP$ algebras, isolates mechanisms that reappear in the $JPZ$ resonant systems:
(i) an invariant tensor correlates the coefficients of a Chern--Simons
action, (ii) an invertible change of translation generator can change the
torsion/curvature interpretation without changing the abstract algebraic
class, and (iii) a more general Riemann--Cartan action may contain independent
couplings that are not tied together by one fixed Chern--Simons algebra.
The Mielke--Baekler model therefore supplies a useful comparison for the
torsional branches discussed in the main text \cite{Mielke:1991nn,Blagojevic:2003vn}.

\subsection{Poincar\'e and (A)dS Chern--Simons gravity}

Consider the algebra generated by $\{J_a,P_a\}$,
\begin{equation}
[J_a,J_b]=\epsilon_{ab}{}^{c}J_c\,,
\qquad
[J_a,P_b]=\epsilon_{ab}{}^{c}P_c\,,
\qquad
[P_a,P_b]=\sigma\,\epsilon_{ab}{}^{c}J_c\,,
\end{equation}
where $\sigma=0,+1,-1$ denotes the Poincar\'e, AdS, and dS cases. With
\begin{equation}
\mathcal A=\omega^aJ_a+\frac1\ell e^aP_a\,,
\end{equation}
the curvature can be written as
\begin{equation}
\mathcal F
=
F_J^aJ_a+F_P^aP_a\,,
\end{equation}
with
\begin{equation}
F_J^a
=
R^a+\frac{\sigma}{2\ell^2}(e\times e)^a\,,
\qquad
F_P^a
=
\frac1\ell T^a\,.
\end{equation}
Introducing the curvature vector
\begin{equation}
\mathbf F_\sigma^a
=
\begin{pmatrix}
F_J^a\\[2mm]
F_P^a
\end{pmatrix},
\end{equation}
the most general invariant bilinear form is represented, in the ordered basis
$(J,P)$, by
\begin{equation}
\Omega_\sigma
=
\begin{pmatrix}
\alpha_0 & \alpha_1\\
\alpha_1 & \sigma\alpha_0
\end{pmatrix}.
\label{eq:two-generator-invariant-matrix}
\end{equation}
The bulk field equations therefore take the same form as in the $(J,P,Z)$ basis analysis,
\begin{equation}
\Omega_\sigma\,\mathbf F_\sigma^a=0\,.
\label{eq:two-generator-matrix-eom}
\end{equation}
The determinant is
\begin{equation}
\det\Omega_\sigma
=
\sigma\alpha_0^2-\alpha_1^2\,.
\end{equation}
Whenever this determinant is non-zero, the variational equations are
equivalent to full gauge flatness,
\begin{equation}
F_J^a=0\,,
\qquad
F_P^a=0\,,
\end{equation}
or explicitly
\begin{equation}
T^a=0\,,
\qquad
R^a+\frac{\sigma}{2\ell^2}(e\times e)^a=0\,.
\end{equation}
These are the standard locally Minkowski, AdS$_3$, and dS$_3$ sectors.

Up to an exact form, the corresponding Chern--Simons Lagrangian is
\begin{align}
\mathcal L_\sigma
={}&
\alpha_0\left(
\CS(\omega)
+\frac{\sigma}{\ell^2}e\cdot T
\right)
\nonumber\\
&+
\alpha_1\left(
\frac{2}{\ell}e\cdot R
+\frac{\sigma}{3\ell^3}e\cdot(e\times e)
\right)\,.
\label{eq:two-generator-cs-lagrangian}
\end{align}
Thus, even in the familiar two-generator examples, the invariant tensor both
selects the admissible combinations of terms and determines whether the
variational equations imply full gauge flatness.

\subsection{General Mielke--Baekler model}
\label{app:MB-JP}

A useful extension of the preceding algebraic picture is obtained by allowing
the commutator of two translations to contain both generators,
\begin{equation}
[J,J]\sim J\,,
\qquad
[J,P]\sim P\,,
\qquad
[P,P]\sim pJ+qP\,.
\label{eq:pq-algebra}
\end{equation}
The standard Poincar\'e and (A)dS cases correspond to $q=0$ and
$p=\hat{Lambda}$. Under an invertible change of the translation generator,
\begin{equation}
P'=aP+bJ\,,
\qquad
a\neq0\,,
\end{equation}
the parameters become
\begin{equation}
q'=aq+2b\,,
\qquad
p'=a^2p-abq-b^2\,,
\end{equation}
while the combination
\begin{equation}
q'^2+4p'
=
a^2\left(q^2+4p\right)
\end{equation}
changes only by a positive square. Its sign therefore distinguishes the AdS, Poincar\'e, and dS algebraic classes. Different values of $(p,q)$ inside a given class can correspond to different choices of translation generator,
and hence to different gravitational realizations of the same abstract
algebra.

The Mielke--Baekler (MB) model provides the dynamical counterpart of this freedom
\cite{Mielke:1991nn,Blagojevic:2003vn}. Its Lagrangian contains four independent couplings,
\begin{equation}
\mathcal L_{\rm MB}
=2a\,e\cdot R(\omega)
-\frac{\hat\Lambda}{3}e\cdot(e\times e)
+\alpha_3\CS(\omega)+\alpha_4e\cdot T\,.
\label{eq:MB-action}
\end{equation}
By comparison, the standard two-generator $JP$ Chern--Simons
action~\eqref{eq:two-generator-cs-lagrangian} has only two
independent couplings and corresponds to the restricted
choice
\begin{equation}
(a,\hat\Lambda,\alpha_3,\alpha_4)
=\left(
\frac{\alpha_1}{\ell},
-\frac{\sigma\alpha_1}{\ell^3},
\alpha_0,\frac{\sigma\alpha_0}{\ell^2}
\right).
\label{eq:MB-CS-subspace}
\end{equation}

Despite the four independent coefficients, variation with respect to $e$ and $\omega$ gives only two field
equations,
\begin{align}
2aR+2\alpha_4T-\hat\Lambda\,e\times e&=0\,,
\\
2\alpha_3R+2aT+\alpha_4e\times e&=0\,.
\end{align}
Thus $R$, $T$, and $e\times e$ are the three geometric two-form building blocks entering two linear relations. For $\Delta_{\rm MB}:=\alpha_3\alpha_4-a^2\neq0$,
these give
\begin{align}
2T&=\tau\,e\times e\,,
&
\tau&=
\frac{\alpha_3\hat\Lambda+a\alpha_4}
{\Delta_{\rm MB}}\,,
\\
2R&=\rho\,e\times e\,,
&
\rho&=
-\frac{a\hat\Lambda+\alpha_4^2}
{\Delta_{\rm MB}}\,.
\end{align}
The constants $\tau$ and $\rho$ determine the torsion and curvature. Indeed, the torsionless connection $\widetilde\omega=\omega-\tfrac{\tau}{2}e$ satisfies
\begin{equation}
D_{\widetilde\omega}e=0\,,
\qquad
2R(\widetilde\omega)
=\left(\rho-\frac{\tau^2}{4}\right)e\times e\,.
\end{equation}
A flat torsionful connection (obtained by setting $\rho=0$) can coexist with a curved metric. This distinction also enters the dynamics of spinning particles \cite{Batlle:2026ims}.

The comparison is useful because MB starts from four terms independently, whereas the $JP$ Chern--Simons actions are constrained by their algebras and invariant tensors. A non-standard basis of the Poincar\'e/(A)dS gauge algebra permitting a Chern--Simons formulation of the full Mielke--Baekler theory was discussed in Ref.~\cite{Geiller:2020edh}, together with its covariant phase space and quasi-local charges.

A complete reconciliation of these various viewpoints, including generalized
generator redefinitions beyond~\eqref{eq:pq-algebra}, the dS/AdS branches, and the double and triple direct-sum structures found in the $JPZ$ family, remains for future work.

\bibliographystyle{unsrt}
\bibliography{references}

\end{document}